\documentclass[twocolumn,floatfix, showpacs]{revtex4}

\usepackage[final]{epsfig}
\usepackage{amsmath}
\begin{document}

\title{Systematics of the charged-hadron $P_T$ spectrum and the nuclear suppression factor
in heavy-ion collisions from $\sqrt{s_{NN}}=200$~GeV to  $\sqrt{s_{NN}}=2.76$~TeV}
 
\author{Thorsten Renk}
\email{thorsten.i.renk@jyu.fi}
\author{Hannu Holopainen}
\email{hannu.l.holopainen@jyu.fi}
\author{Risto Paatelainen}
\email{risto.s.paatelainen@jyu.fi}
\author{Kari J. Eskola}
\email{kari.eskola@phys.jyu.fi}
\affiliation{Department of Physics, P.O. Box 35, FI-40014 University of Jyv\"askyl\"a, Finland}
\affiliation{Helsinki Institute of Physics, P.O. Box 64, FI-00014 University of Helsinki, Finland}


\begin{abstract}
In this paper, our goal is to make a simultaneous analysis of the high- and low-$P_T$ parts of the charged-hadron $P_T$ spectrum measured by the ALICE collaboration \cite{ALICE-RAA} in central Pb-Pb collisions at $\sqrt{s_{NN}}=2.76$~TeV at the Large Hadron Collider (LHC), based on models which have been successfully applied and constrained in Au-Au collisions at Relativistic Heavy Ion Collider (RHIC). For the hydrodynamical modeling with which we obtain the low-$P_T$ spectrum, we have computed the initial conditions based on perturbative QCD (pQCD) minijet production and saturation. The sensitivity of the obtained charged-hadron $P_T$ spectrum on the hydrodynamic model parameters is studied.
For the high-$P_T$ part, we apply a number of parton-medium interaction models, which are tuned to describe the nuclear suppression factor $R_{AA}$  measured at the RHIC in central Au-Au collisions at $\sqrt{s_{NN}}=200$~GeV. We find that the higher kinematic reach of the LHC, manifest in the hardening of the pQCD parton spectral slope, is in principle very efficient in discriminating the various models. However, due to the uncertainties in the p-p baseline, none of the tested models can be firmly ruled out with the present ALICE data. Comparison with the LHC data in this approach shows that the combined hydrodynamic and pQCD+jet quenching  components can reproduce the data well in the whole measured $P_T$ range. 
\end{abstract}
 
\maketitle

\section{Introduction}
\label{sec1}

The very first ALICE physics results from the $\sqrt{s_{NN}}=2.76$~TeV Pb-Pb collisions at the LHC were the elliptic flow \cite{Aamodt:2010pa}, multiplicities \cite{ALICE-M}, as well as the transverse momentum ($P_T$) spectrum of charged particles \cite{ALICE-RAA}. In the measured charged-particle $P_T$ spectrum, which covers the range $P_T<20$~GeV and extends over three orders of magnitude, two quite distinct regions can be seen for the centralmost collisions: 
at $P_T\lesssim 4-5$~GeV the spectrum is exponential, while at $P_T\gtrsim 5$~GeV it shows a power-law like behaviour. Both of these carry very important information on the QCD dynamics of the system: the low-$P_T$ spectrum, responsible for the hadronic bulk multiplicity, reflects the collective transverse motion (flow) developed in the system during its entire spacetime evolution, while the high-$P_T$ spectrum, and its suppression relative to the yield in proton-proton collisions, tells us about energy losses of high transverse momentum ($p_T$) partons on their way out of the dense bulk matter.

In this paper, our goal is to make a simultaneous analysis of the high- and low-$P_T$ parts of the charged-hadron $P_T$ spectrum measured by the ALICE collaboration \cite{ALICE-RAA} in Pb-Pb collisions, based on models which have been successfully applied and constrained in Au-Au collisions at RHIC. While the extrapolation of the pQCD parts from $\sqrt{s_{NN}}=200$~GeV to  2.76~TeV is straightforward, the nonperturbative features in bulk dynamics and its connection to partonic energy loss contain considerable uncertainties.

A generally accepted framework for describing the bulk QCD-matter evolution in $A$-$A$ collisions, is relativistic hydrodynamics, which at the same time is the only framework which can incorporate the effects of the phase transition predicted by lattice QCD. There has been considerable progress in the hydrodynamical modeling of ultrarelativistic heavy-ion collisions over the last years, and into many directions, too: one has moved from solving
1+2 dimensional boost symmetric ideal hydrodynamical equations \cite{Kolb:2000sd} 
to genuinely 
1+3 D ideal hydrodynamics \cite{Hirano:2001eu,Nonaka:2006yn,Schenke:2010nt}, and to 
1+2 D \cite{Heinz:2005bw,Song:2007ux,Dusling:2007gi,Romatschke:2007mq,Niemi:2011ix} and even 
1+3 D dissipative hydrodynamics \cite{Schenke:2010rr}. 
Furthermore, various hybrid models have recently been developed, where the hydrodynamical evolution is coupled with a hadronic cascade afterburner \cite{Teaney:2000cw,Teaney:2001av,Hirano:2005xf,Nonaka:2006yn,Song:2011hk}. Also genuine event-by-event hydrodynamic studies have been performed, both in the ideal hydrodynamics \cite{Andrade:2006yh,Andrade:2008xh,Petersen:2009vx,Werner:2010aa,Holopainen:2010gz} as well as in the viscous case \cite{Schenke:2010rr}.

Hydrodynamics itself does not provide any prescription for the extrapolation in $\sqrt{s_{NN}}$. The $\sqrt{s_{NN}}$ dependence of bulk dynamics enters mainly via the initial conditions (IC) which are crucial element in these studies. One can try to fit the IC using the available data (multiplicities, $P_T$ spectra, elliptic flow) as constraints. Then, however, the more fitting is done the more one looses in predictive power.
To improve upon this, the IC may be computable using a dynamical model for primary production of QCD quanta, which provides the needed $\sqrt{s_{NN}}$ dependence. 
One such initial-state model, which we will employ in this paper, is the pQCD + saturation approach, the "EKRT model" \cite{Eskola:1999fc}. When combined with ideal hydrodynamics, this model correctly predicted the charged-hadron multiplicities in $\sqrt{s_{NN}}=56$, 130 and 200 GeV Au-Au collisions at RHIC, and within 7 \% also at the LHC \cite{Eskola:2001bf}. Also the centrality dependence of the multiplicity is consistent with the data, see Fig.~23(a) in \cite{:2008ez} and Fig.~4 in \cite{Eskola:2000xq}, and the measured $P_T$ spectra of the bulk of the identified hadrons at RHIC have been reproduced well in this framework \cite{Eskola:2005ue}.  Elliptic flow  has also been successfully addressed \cite{Niemi:2008ta,Kolb:2001qz}, and in particular, the similarity of the differential elliptic flow at RHIC and LHC as well as the increase of integrated elliptic flow from RHIC to LHC was predicted in \cite{Niemi:2008ta}. Furthermore, the emergence of the pQCD tail from the hydrodynamic spectrum at $P_T\sim 4...5$~GeV in central Pb-Pb collisions at the LHC was predicted in \cite{Eskola:2005ue}.

High-$p_T$ partons are created in hard pQCD subprocesses along with bulk multiplicity production in ultrarelativistic heavy-ion (A-A) collisions. The idea that the final state interaction of such partons with the surrounding matter reflects properties of the bulk and that hence a measurement of high $P_T$ hadrons can be used as a probe of the QCD medium, is known as "jet tomography" \cite{Jet1,Jet2,Jet3,Jet4,Jet5,Jet6}. One of the expected signatures of this final state interaction is the suppression of high $P_T$ hadron production in A-A collisions as compared to the scaled expectation from proton-proton (p-p) collisions, measured by the nuclear suppression factor $R_{AA}$. This phenomenon is often called "jet quenching" (although this is is slightly misleading since the observable is not a jet of hadrons but inclusive single-hadron spectrum). Experimentally, jet quenching has been measured at RHIC with great precision as a function of collision centrality and orientation with respect to the reaction plane \cite{PHENIX-RAA-RP}.

A number of parton-medium interaction models has been proposed and tested against the experimental results for $R_{AA}$ together with a well-constrained hydrodynamical description of the bulk medium \cite{JetHyd1,JetHyd2,JetHyd3}. However in systematical comparisons of models with the data, even including observables such as high $P_T$ back to back correlations, ambiguities remain \cite{SysJet1,SysJet2,SysJet3} which do not allow firm answers as to what the correct dynamics of parton-medium interaction is in nature. A partial answer was obtained with regard to models proposing elastic (or more general incoherent) energy loss of leading partons \cite{Elastic1,Elastic2,Elastic3,Elastic4,Elastic5}. Such models fail to reproduce pathlength dependent observables such as the spread between in-plane and out of plane emission of hard hadrons \cite{ElRP1,ElRP2}. This result establishes that quantum coherence is a crucial ingredient of the dynamics, but fails to discriminate among the various QCD radiative energy loss models to the degree that a quantitative extraction of medium parameters would be possible with good accuracy \cite{qhat}.

The underlying reason for this failure is that the nuclear suppression factor is largely independent of the functional form of the energy loss probability distribution $P(\Delta E)$ for a given leading parton \cite{gamma-h}. While different models predict different forms for $P(\Delta E)$, the pQCD parton spectrum at RHIC kinematics is so steeply falling that even a small shift in parton energy due to energy loss to the medium effectively acts like a complete suppression of the parton. Thus, only a small part of $P(\Delta E)$ close to zero energy loss is actually probed in observables.  The vastly larger kinematic range of the LHC, leading to a significantly harder pQCD parton spectrum is expected to change this situation and to allow to probe more deeply into $P(\Delta E)$.

While leading-parton energy loss models are sufficient to describe single inclusive high $P_T$ hadron production, there is a second class of models which treat the whole medium-modification of a parton shower in the medium \cite{JEWEL,YaJEM1,YaJEM2,Q-PYTHIA,Martini} with the aim of eventually describing fully reconstructed jets in heavy-ion collisions. These models are often Monte-Carlo (MC) codes extending vacuum shower codes such as PYTHIA \cite{PYTHIA} or HERWIG \cite{HERWIG} to include the interaction with a medium. A systematic comparison of these codes with pathlength dependent observables is so far largely absent.

In this paper, we make a simultaneous analysis of the high- and low-$P_T$ parts of the charged-hadron $P_T$ spectrum measured by the ALICE collaboration \cite{ALICE-RAA} in $\sqrt{s_{NN}}=2.76$~TeV central Pb-Pb collisions at the LHC. We first study the compatibility of the pQCD+saturation+hydrodynamics framework with the measured low-$P_T$ spectrum, systematically charting the model uncertainties. Then for the high-$P_T$ part, using the obtained hydrodynamical evolution of the bulk matter as background, we investigate what discriminating power the first measurement of $R_{AA}$ at the LHC \cite{ALICE-RAA} offers for models which are tuned to describe the observed nuclear suppression at RHIC. We consider both leading-parton radiation \cite{QuenchingWeights} and elastic \cite{ElRP1} energy losses as well as showers \cite{YaJEM1,YaJEM2} modified by the QCD medium. For other recent works discussing $R_{AA}$ at the LHC, see Refs.~\cite{Che:2011vt,Majumder:2011uk}.

For the $R_{AA}$ study, our strategy is as follows: 
First, in moving from RHIC to LHC using our default hydrodynamical set-up, we compute $R_{AA}$ for different models of parton-medium interaction without any re-tuning of parameters (straight extrapolation). Since some amount of uncertainty is expected to originate from the uncertainties in the hydrodynamical initial state, and due to the fact that energy-loss model parameters are known to differ for the same parton-medium interaction model even among constrained hydrodynamical models \cite{SysJet3}, we re-tune in a second run the model parameters to the best fit of the $\sqrt{s_{NN}}=2.76$~TeV data and quote the difference to the tune at $\sqrt{s_{NN}}=200$~GeV. This allows to gauge how the change in $\sqrt{s_{NN}}$ acts to constrain models. 
Using the best fits which we obtain for the low-$P_T$ and high-$P_T$ parts, we investigate to what extent the 
measured charged-hadron $P_T$ spectrum can be reproduced.

\section{The hydrodynamical bulk description}

\subsection{pQCD+saturation+hydrodynamics framework}
For obtaining the bulk hadron $p_T$ spectra in a hydrodynamical framework, we need to define the QCD matter initial conditions, solve the hydrodynamical equations numerically with a given Equation of State (EoS), compute the thermal particle spectra at freeze-out, and account for the strong and electromagnetic decays of unstable particles. 

We compute the QCD matter initial densities using the EKRT saturation model \cite{Eskola:1999fc}, which is based on collinearly factorized pQCD minijet production and the conjecture of gluon saturation in the transverse plane. In this model, which has been quite successful in predicting the multiplicities from RHIC to LHC,
saturation of primary parton (gluon) production is assumed to take place when the minijet production vertices, each of which occupies a geometric uncertainty area $\sim \pi/p_T^2$, start to overlap. For central $A$-$A$ collisions studied here, the following criterion is fulfilled at saturation:
\begin{equation}
  N(p_0) \pi/p_0^2 = \pi R_A^2, 
  \label{saturation}
\end{equation}
where the number of produced minijets at $p_T\ge p_0$ in the mid-rapidity unit $\Delta y=1$ can be written in terms of the standard nuclear overlap function $T_{AA}$  and leading-order (LO) pQCD cross sections as 
$N(p_0) = T_{AA}({\bf 0})\sigma \langle N \rangle$,
with (see Ref. \cite{Eskola:2005ue} for details) 
\begin{eqnarray}
\nonumber
\sigma \langle N \rangle &=& \int_{p_0^2}dp_T^2
\bigg[ \int_{\Delta y}dy_1\int dy_2 + \int dy_1\int_{\Delta y} dy_2 \bigg] \\
&& \sum_{\langle kl\rangle}\frac{1}{1+\delta_{kl}}
\frac{d\sigma^{AA\rightarrow kl+X}}{dp_T^2dy_1dy_2}.
\end{eqnarray}
Above, $T_{AA}({\bf b}) = \int d^2 {\bf s}\, T_A({\bf s}) T_A({\bf s}-{\bf b})$, where 
$T_A(r)=\int dz n_A(r)$ is the nuclear thickness function computed from the Woods-Saxon nuclear densities $n_A$ with $n_0=0.17$~fm$^{-1}$, $d=0.54$~fm. The inclusive cross section for producing partons of flavours $k$ and $l$ above is, as usual,
\begin{equation}
\frac{d\sigma^{AA\rightarrow kl+X}}{dp_T^2dy_1dy_2}
= K \sum_{ij} x_1 f_{i/A}(x_1,Q^2) x_2f_{j/A}(x_2,Q^2) \frac{d\hat\sigma^{ij\rightarrow kl}}{d\hat t}.
  \label{LOminijets}
\end{equation}
For the nuclear parton distribution functions (nPDFs), we use the CTEQ6L1 PDFs \cite{Pumplin:2002vw}
together with the EPS09 nuclear effects \cite{Eskola:2009uj}. For transverse energy ($E_T$) production, the pQCD minijet calculation can be extended to next-to-leading order (NLO) in an infra-red and collinear singularity safe manner (whereas the number of produced minijets is not well defined beyond LO) \cite{Eskola:2000my}. The corresponding updated $K$ factors are, however, not yet available \cite{EHPT_soon}, which is why we simply perform a LO pQCD calculation here and fit the $K$ parameter in Eq.~(\ref{LOminijets}) so that the minijet production at saturation leads to (after hydrodynamic evolution and resonance decays) the measured charged hadron multiplicity in the LHC Pb-Pb collisions at $\sqrt{s_{NN}}=2.76$~TeV in the 0-5\% centrality class, $dN_{\rm ch}/d\eta = 1584$ \cite{ALICE-M}. The obtained $K$ thus also accounts for all higher-order contributions. 
The centrality selection is simulated by considering a central $A_{\rm eff}$-$A_{\rm eff}$ collision of an effective nucleus $A_{\rm eff} = 192 < A$, as explained in Ref. \cite{Eskola:2001bf}.

Most importantly, with the EKRT-based modeling above, we can also estimate the formation time of the system from the pQCD dynamics: according to the uncertainty principle, $\tau_i\approx 1/p_{\text{sat}}$, where $p_{\text{sat}}$ is the solution of Eq.~(\ref{saturation}). With $K=1.54$, obtained by an iterative fit to the measured multiplicity, we have $p_{\text{sat}}=1.58$~GeV and $\tau_i=0.12$~fm.

In the EKRT framework, where the computation of the minijet $E_T$ production is possible in NLO, the initial conditions should be given in terms of the energy density instead of the entropy density. Once the saturation scale is known, we compute the local initial energy density by distributing the minijet $E_T$ over the transverse plane
at the time $\tau_i$ according to (for more details, see \cite{Eskola:2005ue})
\begin{equation}
  \epsilon(r,\tau_i) 
  = T_A(r)T_A(r) \frac{\sigma\langle E_T\rangle}{\tau_i\Delta y},
  \label{eBC}
\end{equation}
where 
\begin{equation}
\sigma\langle E_T\rangle = \int_{p_0^2} dp_T^2 p_T \frac{d(\sigma\langle N \rangle)}{dp_T^2}.
\end{equation}
Above, we distribute the energy density into the transverse plane according to the binary collision profile, thus our default set-up is the "eBC" initialization. It should be emphasized, however, that since we do not attempt to make the saturation condition (\ref{saturation}) strictly local in the transverse plane (which would lead to a varying $p_{\rm sat}$, see \cite{Eskola:2000xq,Kolb:2001qz} for such discussion), the transverse profile is not uniquely fixed here. 

We use ideal hydrodynamics to describe the space-time evolution of the bulk matter. Since we will consider only mid-rapidity observables here, longitudinal boost invariance as well as neglecting the net-baryon number are very valid approximations. We solve the 2+1 dimensional relativistic hydrodynamical equations,  $\partial_\mu T^{\mu\nu} = 0$, using the SHASTA algorithm \cite{Boris,Zalesak}. For the EoS which closes the set of dynamical equations, we choose the recently developed EoS s95p-v1 from Ref.~\cite{Huovinen:2009yb}. We assume a very rapid thermalization here, taking the formation time $\tau_i$ as the starting time $\tau_0$ for the hydrodynamical evolution. Furthermore, a full chemical equilibrium and zero initial transverse flow are assumed. 

Regarding these initial conditions, we should emphasize three points here. 
First, the early initialization of the hydrodynamical evolution, $\tau_0\propto 1/p_{\rm sat}$, is quite essential, since, as pointed out long ago \cite{Eskola:1988yh}, pQCD minijets do produce a large amount of $E_T$, and only by doing $PdV$ work over a long enough time early on, the hydrodynamically evolving system can sufficiently degrade its transverse energy: As shown in \cite{Eskola:2001bf}, the amount of the measured final-state $E_T$ is only a third of the initially produced $E_T$. 
Second, although the system may not be fully thermal at early times, the early start accounts for the buildup of flow and pressure as well as $PdV$ work during the thermalization stage. 
Third, as discussed e.g. in \cite{Chatterjee:2008tp}, thermal photon production is very sensitive to the hydro initialization time. In order to explain the photon production measured at $p_T\sim 2$ GeV in $\sqrt{s_{NN}}=200$~GeV Au-Au collisions, one needs a substantial thermal photon production component from the QGP. This can be obtained only if the initialization time is small enough.
For these reasons, we believe that to start the hydrodynamical evolution at an early time with zero transverse flow 
is physically a well motivated approximation.

Finally, hadron spectra in the hydrodynamical model are calculated from a constant temperature freeze-out hypersurface
using the Cooper-Frye formula \cite{Cooper}, and accounting for the strong and electromagnetic decays of unstable particles. The freeze-out temperature  $T_F$ is fixed so that we can describe the measured positive pion spectra in $\sqrt{s_{NN}} = 200$ GeV Au-Au collisions at RHIC. The value of $T_F$ is found to be 165 MeV for the eBC profile. 
As shown in \cite{Eskola:2007zc}, for computing the $P_T$ spectra of hadrons, keeping the $T_F$ unchanged from RHIC to LHC is a good approximation to a more dynamical decoupling treatment where the scattering rates are compared with the expansion rate of the system. 

\subsection{The results: Hydrodynamical $p_T$ spectra and their systematics}

\begin{figure}[htb]
\begin{center}
\epsfig{file=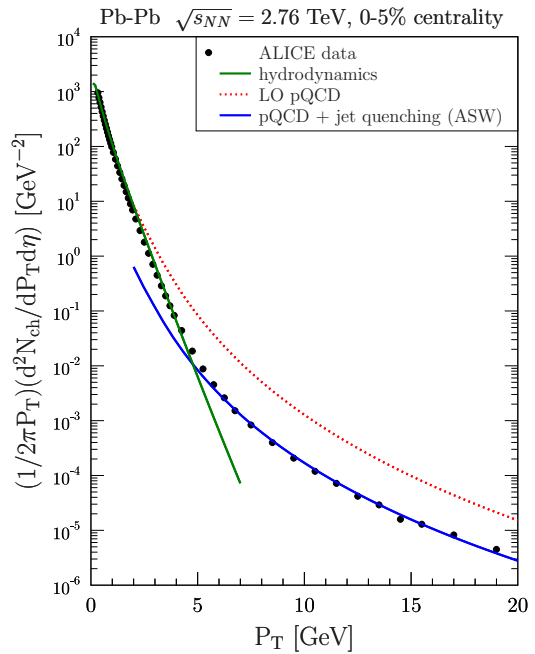, width=8.5cm}
\end{center}
\caption{\label{F-Spectra}(Color online) The transverse momentum $P_T$ spectrum of charged hadrons in 0-5\% central $\sqrt{s_{NN}}=2.76$~TeV Pb-Pb collisions as measured by the ALICE collaboration \cite{ALICE-RAA} and compared with theoretical calculations using a two component picture: the low $P_T$ region is described by pQCD+saturation+hydrodynamics whereas in the high $P_T$ region we apply a pQCD + jet quenching picture, here in the ASW formalism (see text). Shown for comparison is also the pQCD result without jet quenching. Due to the $P_T$ dependence of the jet quenching, this has a different shape which would not agree with the data.}
\end{figure}

Figure~\ref{F-Spectra} shows the hydrodynamically obtained $p_T$ spectrum of charged hadrons in 0-5\% central $\sqrt{s_{NN}}= 2.76$~TeV Pb-Pb collisions at the LHC, and its comparison with the ALICE data \cite{ALICE-RAA}. Also shown is the comparison with the computed pQCD-part of the spectrum, which is subjected to the energy losses of high-energy partons discussed in Sec.~\ref{sec:Elosses} below. As seen in the figure, the agreement of the hydrodynamically obtained spectrum is quite good down to 4-5 GeV, where the pQCD tail takes over. The observed behaviour, the change from the hydro-dominated to the pQCD-dominated spectrum at $p_T= 4-5$ GeV is indeed very similar to what was predicted in \cite{Eskola:2005ue} (see Fig. 15 there).

\begin{figure*}[htb]
\hspace{-.7cm}
\epsfig{file=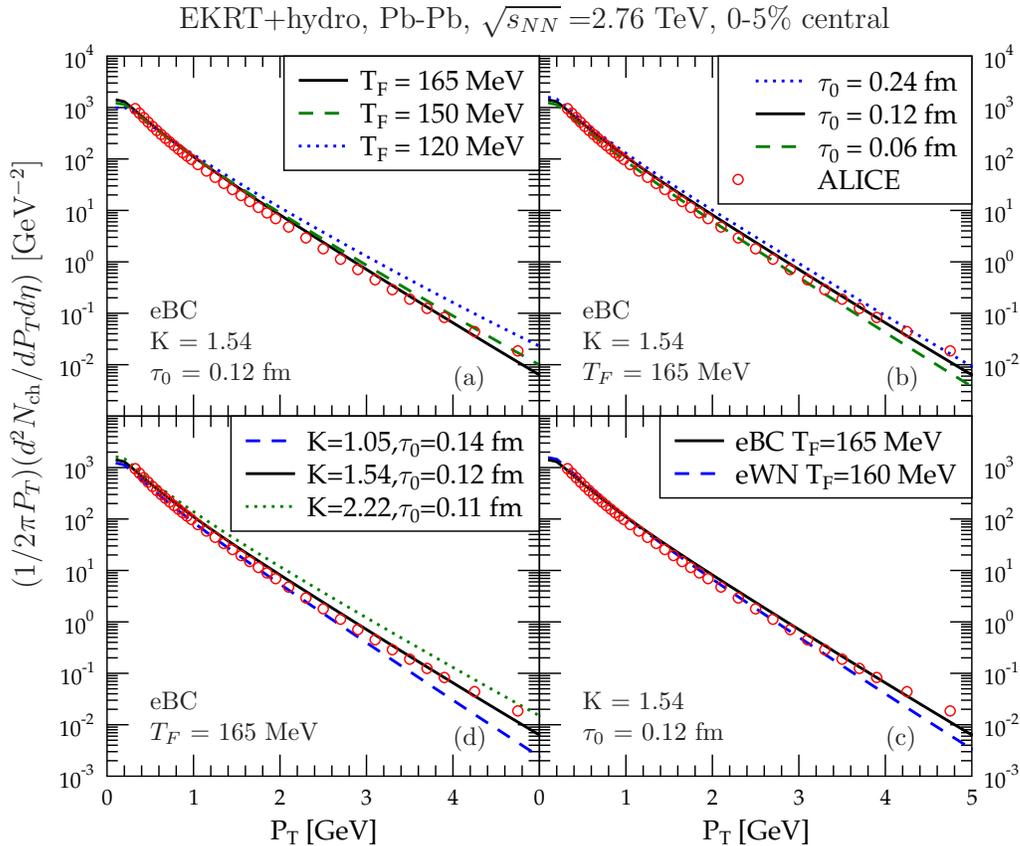, width=14.5cm}
\caption{\label{Hydrosystematics}(Color online) Systematics of the hydrodynamic $P_T$ spectrum in 0-5\% central $\sqrt{s_{NN}}=2.76$~TeV Pb-Pb collisions. Keeping the initial energy fixed, we show the sensitivity of the computed spectrum to 
the freeze-out temperature $T_F$ (upper left panel), to the hydrodynamic initialization time $\tau_0$ (upper right),
and to the initial transverse profile (lower right). Allowing for a change of 20\% for the computed multiplicity, the sensitivity of the results to the value of $K$ (affecting both the initial $E_T$ and $\tau_0$) is shown in the lower left panel. The data points are from the ALICE measurement \cite{ALICE-RAA}.
}
\end{figure*}

To study the sensitivity of the hydrodynamic $p_T$ spectrum to the model uncertainties, we perform the following systematics shown in Fig.~\ref{Hydrosystematics}. 

First, in the upper left panel, we vary the freeze-out temperature between 120 MeV and 165 MeV, keeping the initial conditions fixed to the default set-up. With the single-$T_F$ scenario we use here (i.e. no partial chemical equilibrium or coupling to a hadronic afterburner), a lower $T_F$ enables more transverse flow to develop which in turn leads to a flatter $P_T$ spectrum which easily overshoots the data both at RHIC and at the LHC. Thus, with the eBC and small-$\tau_0$ set-up we need a high value of $T_F$.

Second, the upper right panel shows the sensitivity of the results to the initial time: here $\tau_0$ has been varied in the range $0.5/p_{\rm sat}$ --  $2/p_{\rm sat}$, keeping the initial minijet $E_T$ and the final $T_F$ unchanged. We observe that the eBC set up favours a small hydro initialization time but that at the same time -- since the initial $E_T$ is conserved and not the initial entropy ($S_0$) -- the multiplicity decreases slightly with decreasing $\tau_0$:  
$N_{\rm ch}\sim S_0 \sim \epsilon^{3/4}\tau_0\sim (E_T/\tau_0)^{3/4}\tau_0 \sim E_T^{3/4}\tau_0^{1/4}$. 
Thus, changing $\tau_0$ by a factor 2 causes a change of 19 \% in the multiplicity -- a change which is already well beyond the 5 \% error bar in the data \cite{ALICE-M}. 

Third,  the panel on lower right shows the sensitivity of the hadronic $P_T$ spectrum to the choice of the energy density transverse profile. In addition to the binary profile $T_A(r)T_A(r)$ in Eq.~(\ref{eBC}), we have distributed the minijet $E_T$ also over a wounded nucleon density profile (eWN), keeping however $\tau_0$ fixed. In this change, the charged particle multiplicity then increases from 1580 to 1640. Based on fitting the measured charged-particle $P_T$ spectrum in $\sqrt{s_{NN}}= 200$~GeV Au-Au collisions at RHIC, we use $T_F=160$~MeV for the eWN case. Due to the slower build-up of transverse flow in the eWN case, the eWN spectrum becomes slightly steeper than our default eBC case with $T_F=165$ MeV. From this, we can deduce that fine-tuning the transverse profile towards an eWN profile would be certainly possible but also that an early initial time $\tau_0\approx 1/p_{\rm sat}$ is still required. 

Fourth, the lower left panel shows the sensitivity of the spectrum to the fit parameter $K$. For this figure, we allow a change of 20 \% (ca. 300) in the multiplicity, and compute the $E_T$ and $\tau_0$ separately in each case but keep $T_F$ constant. We can see how an increase(decrease) in the multiplicity corresponds to a larger increase(decrease) in the value of $K$ but a smaller decrease(increase) in $\tau_0$. These scalings can be deduced from Eqs.~(1)--(5), keeping track of the powers of $p_0$ and $K$: At the scaling limit $\sigma(p_0)\sim K/p_0^2$, and Eq.~(\ref{saturation}) gives $p_{\rm sat}\sim K^{1/4}$. The energy density scales as $\epsilon\sim \sigma\langle E_T\rangle p_{\rm sat}
 \sim (K/p_{\rm sat})p_{\rm sat}\sim K$, and the charged-particle multiplicity then scales as $N_{\rm ch}\sim S_0 \sim \epsilon^{3/4}\tau_0\sim K^{1/2}$. Thus, a 20\% increase in the multiplicity corresponds to a 40 \% increase in the value of $K$ and a 10 \% decrease in $\tau_0$. We see on the one hand, that our initial-state modeling is fairly robust against small variations in $K$ and on the other hand that fine-tuning of $K$ would depend on the transverse profile and also on $T_F$. Such fine-tuning, however, we do not consider very meaningful before more data, on e.g. identified hadron spectra, are available. 

From Fig.~\ref{Hydrosystematics}, we conclude on the one hand that with our pQCD+saturation+(ideal)hydrodynamics framework we are committed to a fairly narrow window of the parameters $\tau_0$ and $T_F$, and on the other hand that the profile uncertainty can be considered to be the main uncertainty from the jet quenching viewpoint. Further fine-tuning of the hydrodynamical description is left as future work, when we are studying the centrality dependence of the hadronic $P_T$ spectrum.

\section{Parton-medium interaction models}
\label{sec:Elosses}

The starting point for a computation of the high $P_T$ hadron yield in an A-A collision is the initial spectrum of hard partons.  
In our framework, the differential cross section $d\sigma_{vac}^{AA \rightarrow f +X}$ for the production of a parton $f$ in an A-A collision is calculated in LO pQCD by integrating out the unobserved parton kinematics in Eq.~(\ref{LOminijets}) (explicit expressions are given in \cite{SysJet1} and references therein). 

Uncertainty relation arguments indicate that the medium cannot modify the hard process itself, but rather influences the fragmentation pattern of the outgoing highly virtual partons, in particular their development into a parton shower. Thus, for in-medium shower models, the expression to evaluate is the convolution of the partonic production cross section with the medium-modified fragmentation function (MMFF), 
\begin{equation}
\label{E-Conv}
d\sigma_{\rm med}^{AA\rightarrow h+X} \negthickspace \negthickspace = \sum_f d\sigma_{vac}^{AA \rightarrow f +X} \otimes \langle D_{MM}^{f \rightarrow h}(z,\mu^2)\rangle_{T_{AA}}
\end{equation} 
where $\langle D_{MM}^{f \rightarrow h}(z,\mu^2)\rangle_{T_{AA}}$ is the MMFF, averaged over the geometry of the medium, $z$ is the  fractional momentum of produced hadrons given a parton $f$, and $\mu^2$ is the hadronic momentum scale. To evaluate this expression requires knowledge of both the geometry of the medium (e.g. in terms of a spacetime description of medium density) and of the MMFF $D_{MM}(z, \mu_p^2,\zeta)$ for any given path $\zeta$ through the medium.

If the angle between outgoing parton and the reaction plane is $\phi$, the path of a given parton through the medium $\zeta(\tau)$, i.e. its trajectory $\zeta$ as a function of proper medium evolution time $\tau$ is determined in an eikonal approximation by its initial position ${\bf r_0}$  and the angle $\phi$ as $\zeta(\tau) = \left(x_0 + \tau \cos(\phi), y_0 + \tau \sin(\phi)\right)$ where the parton is assumed to move with the speed of light $c=1$ and the $x$-direction is chosen to be in the reaction plane. How $D_{MM}(z, \mu_p^2,\zeta)$ is obtained once a medium is specified is characteristic for a given model of parton-medium interaction and will be discussed for the code YaJEM later.

Once the MMFF for a given path is known, the averaging over the medium geometry is given by 
\begin{equation}
\label{E-D_TAA}
\begin{split}
\langle D_{MM}&(z,\mu_p^2)\rangle_{T_{AA}} \negthickspace =\\ &\negthickspace \frac{1}{2\pi} \int_0^{2\pi}  
\negthickspace \negthickspace \negthickspace d\phi 
\int_{-\infty}^{\infty} \negthickspace \negthickspace \negthickspace \negthickspace dx_0 
\int_{-\infty}^{\infty} \negthickspace \negthickspace \negthickspace \negthickspace dy_0 P(x_0,y_0)  
D_{MM}(z, \mu_p^2,\zeta).
\end{split}
\end{equation}
Here, the initial distribution of hard vertices in the transverse $(x,y)$ plane is assumed to be calculable  as
\begin{equation}
\label{E-Profile}
P(x_0,y_0) = \frac{T_{A}({\bf r_0 + b/2}) T_A(\bf r_0 - b/2)}{T_{AA}({\bf b})}.
\end{equation}

In the leading-parton energy loss approximation, the medium-modified production of high-$P_T$ hadrons can be computed from the convolution
\begin{equation}
d\sigma_{\rm med}^{AA\rightarrow h+X} \negthickspace \negthickspace = \sum_f d\sigma_{\rm vac}^{AA \rightarrow f +X} \otimes \langle P(\Delta E)\rangle_{T_{AA}} \otimes
D^{f \rightarrow h}(z, \mu^2)
\end{equation} 
where $\langle P(\Delta E)\rangle_{T_{AA}}$ is the medium-induced energy loss probability, averaged over the medium geometry and $D^{f \rightarrow h}(z, \mu^2)$ is the vacuum fragmentation function for the production of a hadron $h$ from a parton $f$, at fractional momentum $z$ and hadronic momentum scale $\mu^2$. The underlying assumption is that the dynamics of parton-medium interactions can largely be cast in terms of a shift in leading parton energy and that hence the MMFF can be approximated by the convolution of an energy loss probability with the vacuum fragmentation function, $\langle D_{MM}^{f \rightarrow h}(z,\mu^2)\rangle_{T_{AA}} = \langle P(\Delta E)\rangle_{T_{AA}} \otimes D^{f \rightarrow h}(z, \mu^2)$. In this case, the energy-loss probability for a given path $\zeta$ of a parton through the medium, $P(\Delta E, \zeta)$, is the ingredient to be computed within a specific model of parton-medium interaction.

If $P(\Delta E, \zeta)$ is known, the geometrical averaging involves as above integrating over all possible initial vertices $(x_0,y_0)$ with the weight of $P(x_0,y_0)$ and over all possible orientations $\phi$ as
\begin{equation}
\label{E-P_TAA}
\langle P(\Delta E)\rangle_{T_{AA}} \negthickspace = \negthickspace \frac{1}{2\pi} \int_0^{2\pi}  
\negthickspace \negthickspace \negthickspace d\phi 
\int_{-\infty}^{\infty} \negthickspace \negthickspace \negthickspace \negthickspace dx_0 
\int_{-\infty}^{\infty} \negthickspace \negthickspace \negthickspace \negthickspace dy_0 P(x_0,y_0)  
P(\Delta E,\zeta).
\end{equation}

In all cases, the nuclear modification factor is computed with the given medium-modified yield of hard hadron production as
\begin{equation}
\label{E-RAA}
R_{AA}(p_T,y) = \frac{dN^h_{AA}/dp_Tdy }{T_{AA}({\bf b}) d\sigma^{pp}/dp_Tdy}.
\end{equation}

The details of the parton-medium interaction model are thus in either the energy loss probability distribution $P(\Delta E,\zeta)$ for leading parton energy loss models or the MMFF $D_{MM}(z, \mu_p^2,\zeta)$ for in-medium shower models, given a specific path through the medium. In the following, we formulate three different types of models which we apply to the ALICE data.

\subsection{Armesto-Salgado-Wiedemann (ASW) formalism}

The detailed calculation of $P(\Delta E, \zeta)$ follows the Baier-Dokshitzer-Mueller-Peigne-Schiff (BDMPS) formalism for radiative energy loss  \cite{Jet2} using quenching weights as introduced by Salgado and Wiedemann \cite{QuenchingWeights}.

In this framework, the energy loss probability $P(\Delta E,\zeta)$ for a path can be obtained by evaluating the line integrals along $\zeta(\tau)$ as
\begin{equation}
\label{E-omega}
\omega_c({\bf r_0}, \phi) = \int_0^\infty \negthickspace d \zeta \zeta \hat{q}(\zeta) \quad  \text{and} \quad \langle\hat{q}L\rangle ({\bf r_0}, \phi) = \int_0^\infty \negthickspace d \zeta \hat{q}(\zeta)
\end{equation}
with the relation 
\begin{equation}
\label{E-qhat}
\hat{q}(\zeta) = K_{\rm med} \cdot 2 \cdot \epsilon^{3/4}(\zeta) (\cosh \rho - \sinh \rho \cos\alpha)
\end{equation}
assumed between the local transport coefficient $\hat{q}(\zeta)$ (specifying the quenching power of the medium), the energy density $\epsilon$ and the local flow rapidity $\rho$ with angle $\alpha$ between flow and parton trajectory \cite{Flow1,Flow2}. $K_{\rm med}$ is the adjustable parameter in this framework. It is naturally expected to be $O(1)$, but in fits to data at $\sqrt{s_{NN}}=200$~GeV the parameter takes (dependent on the precise hydrodynamical model) values ranging between 3 and 10 (the latter number occurs for viscous hydrodynamics where the initial entropy density is lower than in the ideal case, see \cite{SysJet3}).

Using the numerical results of \cite{QuenchingWeights} and the definitions above, the energy loss probability distribution given a parton trajectory can now be obtained as a function of the initial vertex and direction $({\bf r_0},\phi)$ as $P(\Delta E; \omega_c({\bf r},\phi), R({\bf r},\phi)) \equiv P(\Delta E,\zeta)$ for $\omega_c(\zeta)$ and $R=2\omega_c(\zeta)^2/\langle\hat{q}L(\zeta)\rangle$. In practical terms, $\langle P(\Delta E) \rangle_{T_{AA}}$ is characterized by a fairly large discrete escape probability without energy loss and a very broad distribution of energy loss ranging up to $O(100)$ GeV at RHIC conditions (for explicit figures, see e.g. \cite{SysJet1}). 

\subsection{YaJEM (Yet another Jet Energy-loss Model)}

The MC code YaJEM is based on the PYSHOW code \cite{PYSHOW} which is part of PYTHIA \cite{PYTHIA}. It simulates the evolution from a highly virtual initial parton to a shower of partons at lower virtuality in the presence of a medium. A detailed description of the model can be found in \cite{YaJEM1,YaJEM2}.

The parton shower developing from a highly virtual initial hard parton in this model is described as a series of $1\rightarrow 2$ splittings $a \rightarrow bc$ where the virtuality scale decreases in each splitting, i.e. $Q_a > Q_b,Q_c$ and the energy is shared among the daughter partons $b,c$ as $E_b = z E_a$ and $E_c = (1-z) E_a$. The splitting probabilities for a parton $a$ in terms of $Q_a, E_a$ are calculable in pQCD and the resulting shower is computed event by event in a MC framework.  In the presence of a medium, the main assumption of YaJEM is that the parton kinematics or the splitting probability is modified. In the RAD (radiative energy loss) scenario, the relevant modification is a virtuality gain
\begin{equation}
\label{E-Qgain}
\Delta Q_a^2 = \int_{\tau_a^0}^{\tau_a^0 + \tau_a} d\zeta \hat{q}(\zeta)
\end{equation}
through the interaction with the medium during the parton lifetime $\tau_a$. This modification leads to an increase in radiation. In order to evaluate Eq.~(\ref{E-Qgain}) during the shower evolution, the momentum space variables of the shower evolution equations need to be linked with a spacetime position in the medium. This is done via the uncertainty relation for the average formation time as
\begin{equation}
\label{E-Lifetime}
\langle \tau_b \rangle = \frac{E_b}{Q_b^2} - \frac{E_b}{Q_a^2}
\end{equation} 
and randomized splitting by splitting by sampling $\tau_b$ from the distribution
\begin{equation}
\label{E-RLifetime}
P(\tau_b) = \exp\left[- \frac{\tau_b}{\langle \tau_b \rangle}  \right].
\end{equation}

The evolution for any given parton in the shower evolution is terminated as soon as the parton reaches a minimum virtuality scale $Q_0$. The result of the partonic evolution in terms of a shower of low virtuality partons is then passed on to the Lund model \cite{Lund} to hadronize. The fractional longitudinal momentum distribution of the resulting hadron distribution corresponds to the MMFF of the various hadron species.

In the default version of YaJEM, the minimum virtuality scale is fixed at $Q_0 = 0.7$ GeV. In the version YaJEM-D (dynamical computation of $Q_0$) \cite{YaJEM-D}, the formation length of the in-medium shower is forced to be within the medium length. This corresponds to the choice
\begin{equation}
\label{E-Q0}
Q_0 = \sqrt{E/L}
\end{equation}
which depends on both in-medium pathlength $L$ and shower-initiating parton energy $E$. The original motivation for this prescription was to introduce a pathlength dependence that can account for the experimentally observed split between in-plane and out of plane emission of high $P_T$ hadrons in non-central collisions \cite{PHENIX-RAA-RP}. However, together with the stronger pathlength dependence, YaJEM-D also predicts a strong rise of $R_{AA}$ with $P_T$ in angular averaged observables which we aim to test against the ALICE data.

In principle, the full functional form of $\hat{q}(\zeta)$ could determine the MMFF, which would be computationally very expensive as a full MC simulation would be needed for every possible path in the medium. However, due to an approximate scaling law identified in \cite{YaJEM1}, it is sufficient to compute the line integral
\begin{equation}
\label{E-Qsq}
\Delta Q^2_{\rm tot} = \int d \zeta \hat{q}(\zeta)
\end{equation}
in the medium to obtain the full MMFF $D_{MM}(z, \mu_p^2,\zeta)$ from a YaJEM simulation for a given eikonal path of the shower-initiating parton, where $\mu_p^2$ is the  momentum scale of the shower initializing parton. The scaling law implies that the MC simulation has to be run only for a finite set of paths and makes a numerical solution of the geometry averaging possible.

Each YaJEM run determines the MMFF for a fixed partonic scale $\mu_p$. To account for the scale evolution of the MMFF, runs for different $\mu_p$ need to be done. For technical reasons having to do with numerical performance, we like to evolve the MMFF for given hadronic scale as indicated in Eq.~(\ref{E-Conv}). For matching a partonic scale for which the MMFF is computed to a hadronic scale, we use the following procedure:
For each available partonic  scale, $\langle D_{MM}(z,\mu_p^2)\rangle_{T_{AA}}$ is computed, and the exponent $n$ of a power law fit to the parton spectrum at scale $\mu_p$ is determined. The maximum of $z^{n-1} \langle D_{MM}(z,\mu_p^2)\rangle_{T_{AA}}$ corresponds to the most likely value $\tilde{z}$ in the fragmentation process, and thus the partonic scale choice is best for a corresponding hadronic scale $P_T = \tilde{z}\mu_p$. The $P_T$ dependence of the hadronic $R_{AA}$ is then computed by interpolation between runs with different scale choice for the MMFF.

As in the previous case, $K_{\rm med}$ in Eq.(\ref{E-qhat}) serves as the adjustable parameter of the model once $Q_0$ is chosen. YaJEM requires, dependent on the underlying hydrodynamical model, rather natural values for $K_{\rm med}$ ranging from 0.6 to 2. 

\subsection{Parametrized elastic energy loss}

In \cite{ElRP1}, a phenomenological model for elastic energy loss, consisting of a discrete parton escape probability and a Gaussian parametrization for the energy loss probability was introduced to explore the pathlength dependence of incoherent energy loss. While the model itself is rather simplistic, its main findings with regard to pathlength dependence have been confirmed later in a detailed MC simulation of elastic energy loss \cite{ElRP2}. It needs to be stressed that unlike the previous models, the parametrized elastic energy loss is not meant as a serious QCD based model for the underlying dynamics of parton-medium interaction. The reason that it is presented here is rather that the simple and adjustable form of $P(\Delta E,\zeta)$ allows insight into how the observed rise of $R_{AA}$ with $P_T$  depends on the underlying functional form of the energy loss probability density, which is much less transparent in the context of ASW or YaJEM.

In the model, the escape probability of a parton $i$ without any medium interaction is computed as
\begin{equation}
P_{\rm esc}^i = \exp\left[- const. \cdot \sigma_{el}^i \int \tilde{\rho}_M(\zeta)d\zeta \right] = \exp[- \gamma_i \cdot \kappa]
\end{equation}
where it is assumed that $\sigma_{el}$ is approximately independent of $\zeta$, and $\kappa$ is defined  as
\begin{equation}
\label{E-kappa}
\kappa = \int d\xi \epsilon^{3/4}(\xi) (\cosh \rho(\xi) - \sinh \rho(\xi) \cos\alpha)
\end{equation}
taking into account the flow corrections to the probed density. Here $\gamma_i$ is a parameter with dimensions of a cross section measuring the interaction strength, and hence $\gamma_g = 9/4 \gamma_q$ must hold to account for the different color factors of quarks and gluons.

If the parton does not escape without energy loss, it must undergo a shift in energy (there is also the possibility that a strong shift into a thermal regime occurs, which is equivalent to an absorption of the parton). It is assumed that the mean value of the shift in energy will grow linearly in the number of scatterings $N$ as 
\begin{equation}
d\Delta E = \Delta E_{1} \sigma_{el}^i \rho_M d\xi
\end{equation}
with $\Delta E_1$ the mean energy loss per scattering, whereas the fluctuations around the mean will grow like $\sqrt{N}$. Assuming a Gaussian distribution, this leads to the ansatz
\begin{equation}
\label{E-Elastic}
P^i(\Delta E,\zeta) = P_{\rm esc}^i \delta(\Delta E) + \mathcal{N}_i \exp\left[ \frac{(\Delta E - \alpha_i \kappa)^2}{\beta_i \kappa}  \right]
\end{equation}
where $\mathcal{N}_i$ is a normalization such that $\int_0^\infty d (\Delta E) P^i(\Delta E, \zeta) = 1$ and (\ref{E-Elastic}) has to hold for quarks and gluons separately due to the different color factor. $\alpha_i$ is a parameter with the dimensions of a cross section times the energy shift per reaction.

The model is thus characterized by three parameters:

\begin{itemize}
\item $\alpha_i$ controls the mean shift in energy per expected scattering
\item $\beta_i$ governs the strength of fluctuations around this mean shift. If $\beta_i$ is small, the model will have a strong correlation between path (and hence initial vertex) and shift in energy, if the parameter is large, this correlation is lessened
\item $\gamma_i$ finally determines the magnitude of the escape probability.
 \end{itemize}

In \cite{ElRP1} it was discussed that the space of all possible $(\alpha_i, \beta_i, \gamma_i)$ which can describe the measured $R_{AA}$ for $\sqrt{s_{NN}}=200$~GeV central Au-Au collisions is triangular and ordered by $\gamma$ --- if $P_{\rm esc}$ is of the order of the measured $R_{AA}$ already, the space of allowed shifts in energy is very constrained. On the other hand, if $P_{\rm esc}$ is small, more possibilities for a shift arise.

Here, we investigate two distinct scenarios, one with large escape probabilities for quarks and gluons close to the allowed limit where $P_{\rm esc}^q = 0.218, P_{\rm esc}^g = 0.054$ and one with about half these values where $P_{\rm esc}^q = 0.12, P_{\rm esc}^g = 0.027$. The parameters are found in Table \ref{T-Params}. Note that the parameters here do not correspond to the sets given in \cite{ElRP1} since we use a different hydrodynamical model to do the extrapolation from  $\sqrt{s_{NN}}=200$~GeV to 2.76~TeV here.

\begin{table}[htb]
\begin{tabular}{ccc}
\hline
&small $P_{\rm esc}$ & large $P_{\rm esc}$ \\
\hline \hline
$\alpha_q$ [GeV$^{-1}$] & 0.5 &2.0 \\
$\beta_q$ & 20.0 &  80.0\\
$\gamma_q$ [GeV$^{-2}$] & 0.4 & 0.28 \\
\hline
\end{tabular}
\caption{\label{T-Params}Parameters for the two different elastic energy loss scenarios described in the text.}
\end{table}

\section{Results}

\subsection{$P_T$ spectra}

In Fig.~\ref{F-Spectra}, we show in addition to the good description of low $P_T$ bulk matter by hydrodynamics the quality of the description of the high $P_T$ part above 6 GeV by pQCD + jet quenching. We conclude from this comparison that a LO pQCD calculation, supplemented by a $K$ factor, is sufficient for the accuracy required to compute $R_{AA}$ reliably enough. The residual small deviations are not a crucial issue, 
 since due to the nature of $R_{AA}$ as a ratio of $P_T$ spectra deviations in the spectral shape between computation and data cancel to first order and only lead to subleading corrections.

We also note at this point that the effect of jet quenching cannot be cast into the form of a constant downward shift of the unmodified spectrum, but that rather its $P_T$ dependence is crucial to describe the spectral shape correctly. As an interesting side remark, the fact that pQCD and jet quenching offer a good description of the spectrum from 6 to 20 GeV suggests that there is fairly little room for another $P_T$-dependent component of hadron production, such as suggested by, e.g. certain recombination models (see e.g. \cite{Coalescence}) in this region. On the other hand, e.g. the sudden recombination model \cite{Reco,RecoLHC} expects to see effects largely at lower momenta. However, since recombination models generically expect quite different effects for mesons and baryons \cite{Coalescence,Reco,Greco}, the charged hadron spectrum is not a suitable observable to gauge the importance of recombination in this momentum window and a detailed discussion should be based on identified hadron spectra. 

\subsection{Direct extrapolation}

\begin{figure}[htb]
\begin{center}
\epsfig{file=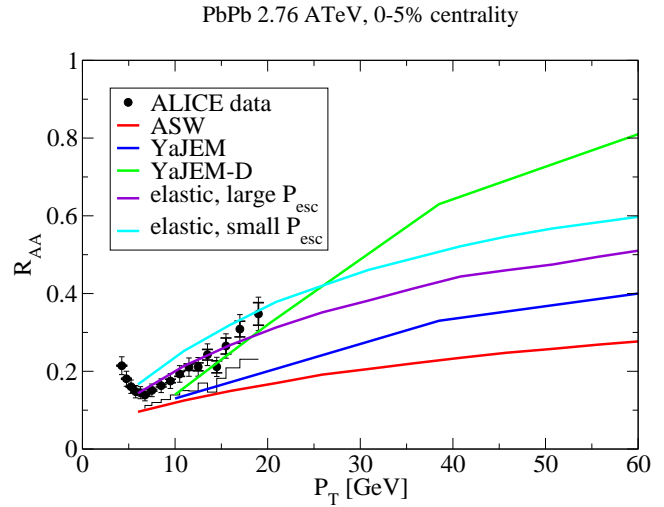, width=8.5cm}
\end{center}
\caption{\label{F-RAA}(Color online) The nuclear suppression factor $R_{AA}$ in 0-5\% central $\sqrt{s_{NN}}=2.76$~TeV Pb-Pb collisions computed in various models for the parton-medium interaction (see text) with model parameters adjusted to describe 0-5\% central  $\sqrt{s_{NN}}=200$~GeV Au-Au collisions compared with the ALICE data \cite{ALICE-RAA} .}
\end{figure}

In Fig.~\ref{F-RAA} we show $R_{AA}$ computed in the various models for parton-medium interaction discussed above in comparison with the ALICE data \cite{ALICE-RAA}, with their parameters adjusted to 0-10\% central $\sqrt{s_{NN}}=200$~GeV Au-Au collisions at RHIC. The assumption underlying this extrapolation is that 
the hydrodynamical model for the bulk matter can be extended from RHIC to LHC in a well controlled manner.
This is a non-trivial issue, as it is known that changing the hydrodynamical model at the same energy may amount to 50\% change in model parameters if the underlying dynamics is sufficiently different, even if both models are constrained by bulk observables. 

The predictions for $R_{AA}$ from the various models at $\sqrt{s_{NN}}=2.76$~TeV turn out to be quite dramatically different in normalization, shape and their expectation for even larger $P_T$. In particular, the differences between models are significantly larger than the statistical errors of the measurement, thus allowing for a clean discrimination if the systematic errors (such as those with the p-p baseline) can be understood.

ASW and YaJEM in the default mode predict a rather slow rise of $R_{AA}$ with $P_T$ along with a strong suppression. This is not in agreement with the data shown for the default p-p reference, but would agree better with an alternative NLO scaled p-p reference (see \cite{ALICE-RAA} for details).

Both parametrized elastic scenarios reproduce the shape of $R_{AA}$ with the default reference well, they mainly differ in the normalization. The difference between the parametrized elastic and the ASW model can be understood as follows: At RHIC conditions, only a narrow region of $P(\Delta E)$ around zero is effectively probed due to the steeply falling spectrum, as  even a small shift in parton energy is equivalent to a massive suppression. In the parametrized elastic models, most of the weight of  $\langle P(\Delta E)\rangle_{T_{AA}}$ is contained in the region between zero and $\sim 30$ GeV (see e.g. Fig.~1 right in \cite{ElRP1}), for small $P_{\rm esc}$ even more weight is contained close to the origin. This is very different for $\langle P(\Delta E)\rangle_{T_{AA}}$ computed in the ASW model --- here (see e.g. Fig.~5 in \cite{SysJet1})  the distribution is much flatter and contains a sizeable weight out to 100 GeV.

To illustrate this in more detail: In a schematic model neglecting (among other things) hadronization, $R_{AA}$ can be understood from the ratio of modified over unmodified parton spectrum, where the modified parton spectrum at a given $p_T$ is determined by the number of partons escaping without energy loss plus the number of partons available in the spectrum at $p_T + \Delta E$ times the probability $P(\Delta E)$ of a shift by $\Delta E$. If we assume a power law $p_T^{-n}$ for the parton spectrum,
\begin{equation}
R_{AA} \approx \int_0^{E_{\rm max}} d\Delta E \langle P(\Delta E) \rangle_{T_{AA}} \left(1+\frac{\Delta E}{p_T}\right)^{-n}.
\end{equation}
It is evident from the expression that $R_{AA}$ at a given $p_T$ is equal to the transmission term of zero energy loss plus a contribution which is proportional to the integral of $\langle P(\Delta E) \rangle_{T_{AA}}$ from zero up to the energy scale $E_{\rm max}$ of the parton (since a parton cannot lose more energy than it originally has), {\it seen through the filter} of the steeply falling spectrum. $R_{AA}$ then generically grows with $p_T$ since $E_{\rm max}$ grows linearly with $p_T$, and the speed of growth depends on the weight of $\langle P(\Delta E) \rangle_{T_{AA}}$ in the region from zero to $E_{\rm max}$ and on the power $n$ of the parton spectrum. 

At the LHC, more of $\langle P(\Delta E) \rangle_{T_{AA}}$ is accessible compared with RHIC due to the harder underlying parton spectrum, i.e. the power $n$ of the 'filter' by which large $\Delta E$ in $\langle P(\Delta E) \rangle_{T_{AA}}$ are suppressed is reduced. This translates into a stronger rise of $R_{AA}$ with $P_T$ \cite{ElRP1} as compared to RHIC kinematics, and this rise is most pronounced for models where there is substantial weight of the energy loss probability density close to the origin. Thus, the parametrized elastic model with small $P_{\rm esc}$ shows the strongest rise, while ASW shows very weak $P_T$ dependence.

Finally, YaJEM-D predicts the strongest rise of $R_{AA}$ with $P_T$ as a consequence of Eq.~(\ref{E-Q0}). In addition to the rise expected from the way the effective energy loss probability is probed as outlined above, YaJEM-D thus contains an explicit mechanism which introduces a strong energy dependence into the MMFF itself. While the normalization of the curve falls below the default baseline data, the shape is well reproduced.

\subsection{Refit to data}

\begin{figure}[htb]
\begin{center}
\epsfig{file=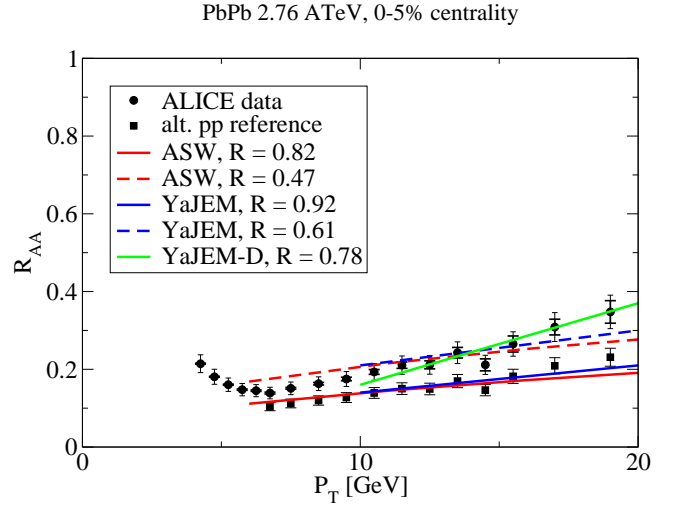, width=8.5cm}
\end{center}
\caption{\label{F-RAArefit}(Color online) The nuclear suppression factor $R_{AA}$ in 0-5\% central $\sqrt{s_{NN}}=2.76$~TeV Pb-Pb collisions computed in various models for the parton-medium interaction (see text) with model parameters refit to  ALICE data \cite{ALICE-RAA}. To indicate the magnitude of the uncertainty, the alternative p-p reference result which is not shown with errors by ALICE has been given a 10\% error band.}
\end{figure}

As discussed in the context of Fig.~\ref{Hydrosystematics} above, there can be residual uncertainties in extrapolating the QCD matter fluid dynamics description from RHIC to the LHC. Therefore, in the second step, we allow a refit of parton-medium interaction model parameters to the ALICE data. We introduce a parameter $R$ to quantify the amount of refitting which is needed, where $R$ stands for the ratio of modified over unmodified parameter $K_{\rm med}$. Here we would consider, say, a 25\% change in the model parameter reasonable within the uncertainties of the hydrodynamical extrapolation. The results are shown in Fig.~\ref{F-RAArefit}.

Following this procedure, we find that the ASW model can be brought into a rough agreement with the alternative p-p reference data, but even allowing for a substantial parameter readjustment it does not agree with the default reference data --- the $P_T$ dependence is too weak.
YaJEM likewise follows the trend of the alternative p-p reference data well. YaJEM-D on the other hand can be brought into good agreement in both shape and normalization with the default p-p reference data.

Unfortunately, even provided that we are willing to discard a model based on deviations $R < 0.75$ or $R > 1.25$  in the refitting procedure, we can at present make no such statement from the ALICE data due to the uncertainty in the p-p baseline. This stresses the importance of having a measured baseline. However, looking at Fig.~\ref{F-RAA}, we see reason to conclude that already a larger range in $P_T$ will allow to decisively rule out some models based on the wrong shape of the $P_T$ dependence. 

\section{Discussion}

In this work, we have simultaneously analyzed the low-$P_T$ and high-$P_T$ spectrum of charged hadrons in 
central Pb-Pb collisions at $\sqrt{s_{NN}}=2.76$~TeV. For hydrodynamical modeling describing the low-$P_T$ spectrum, we have computed the initial conditions based on pQCD minijet production and saturation \cite{Eskola:1999fc,Eskola:2005ue}. To account for the NLO and higher order corrections in minijet production, we have made an iterative fit to the measured LHC charged-hadron multiplicity. The main outcome of this computation are the produced transverse energy and the early formation time, $\tau_i\approx 0.12$~fm, of the pQCD minijet system. We assume that $\tau_i$ is also the starting time $\tau_0$ for the hydrodynamical evolution but in charting the uncertainties of our modeling, we have shown the sensitivity of the computed $P_T$ spectrum to $\tau_0$. Also the sensitivity to the decoupling temperature,  multiplicity fitting, and transverse profile of the initial energy density is explicitly shown. 

In our hydrodynamic framework, the uncertainty related to the initial transverse profile can be considered to be the main uncertainty in the extrapolation from RHIC to LHC. We have for simplicity considered two possibilities here, the eBC and the eWN profiles, thus essentially ignoring the QCD dynamics that may cause the profile to change from RHIC to LHC. The profile should, however, be computable in the pQCD+saturation approach \cite{Eskola:2000xq} but since we do not have the needed NLO pQCD elements fully at hand yet \cite{EHPT_soon}, we leave this as interesting future work. We nevertheless can see that within the uncertainties charted,  the pQCD + saturation + (ideal)hydrodynamics framework works quite well in reproducing the charged-hadron $P_T$ spectrum up to $P_T\sim 4$~GeV, and that we obtain the agreement essentially without further tuning of the hydrodynamic parameters from RHIC to LHC. Also interestingly, we have observed (see Fig.~\ref{F-Spectra}) that once parton energy losses have been accounted for, the computed pQCD tail of high-$P_T$ hadron production  starts to dominate over the hydrodynamic component at $P_T\gtrsim 5$~GeV. Comparison with the LHC data shows that the matching of these two components is very efficient in that it leaves fairly little room for hadron production components in addition to the hydrodynamics and pQCD+energy loss in the cross-over region $P_T=4-5$~GeV.

After getting the obtained hydrodynamical evolution of the background QCD matter in control, we have proceeded to analyze the high-$P_T$ part of the charged-hadron $P_T$ spectrum. 
We have applied several models of parton-medium interactions, tuned to the nuclear suppression factor $R_{AA}$ at $\sqrt{s_{NN}}=200$~GeV Au-Au collisions, to $\sqrt{s_{NN}}=2.76$~TeV Pb-Pb collisions where the primary parton spectra are significantly harder. We found that in principle the large kinematic lever-arm at the LHC translates into a significant power to distinguish various models, even given the uncertainties in extrapolating the bulk medium model to larger energy. Of particular importance here is the rise of $R_{AA}$ with $P_T$ observed at $\sqrt{s_{NN}}=2.76$~TeV which is intimately connected with the slope of the pQCD parton spectrum and reflects the way the energy loss probability distribution is probed.

If one had no systematic uncertainty in the data at this point, two of the models we tested (ASW and YaJEM) could be ruled out already by the ALICE data obtained with the default p-p baseline. On the other hand, these models agree fairly well with the alternative p-p reference data. However, in addition to the angular averaged suppression factor in central collisions considered here, other observables need to be studied. If we require that a model should also account for the observed spread between the in-plane and out-of-plane hard hadron emission as observed at $\sqrt{s_{NN}}=200$~GeV, 
then YaJEM would be strongly disfavoured \cite{YaJEM-D} and YaJEM-D would be more consistent with the data than ASW. 

The fact that the simple parametrized elastic energy loss model which is known to fail for pathlength dependent observables is able to describe the scaling in $\sqrt{s}$ from RHIC to LHC rather well should be a stern warning that $\sqrt{s}$ and $P_T$ dependence only probe particular aspects of energy loss models, and that agreement with a subset of available observables may not be enough to judge the validity of a model.

As shown in Fig.~\ref{F-RAA}, extending the measurements of $R_{AA}$ out to larger values of $P_T$ at the LHC will provide strong constraints and viability tests for the parton-medium interaction models.  A combined analysis of $R_{AA}$ in the dependence on $P_T, \sqrt{s_{NN}}$, impact parameter $b$ and reaction plane angle $\phi$ will be highly discriminating between the available models even without having to resort to other high $P_T$ observables such as triggered correlation measurements or fully reconstructed jets. Constraining the nature of the parton-medium interaction by leading hadron production in this way is thus the first step towards tackling the more difficult task of understanding the complete dynamics of a parton shower in the medium.

\begin{acknowledgments}
We thank Pasi Huovinen for providing us with the EoS we used in this work.
T.R. is supported by the Academy researcher program of the
Academy of Finland, Project No. 130472. Additional support comes from K.J.E's
Academy Project 133005. H.H. gratefully acknowledges the financial
support from the national Graduate School of Particle and Nuclear Physics,
and the computing time from the CSC  IT Center for Science at Espoo, Finland. 
 
\end{acknowledgments}

\end{document}